\documentclass[iop]{emulateapj}
\usepackage{subfigure}
\usepackage{wrapfig}
\usepackage{rotating}

\newcommand{\beq}{\bigskip\begin{equation}}
\newcommand{\eeq}{\bigskip\end{equation}}

\def\lsim{\mathrel{\rlap{\lower4pt\hbox{\hskip1pt$\sim$}}
    \raise1pt\hbox{$<$}}}                % less than or approx. symbol
\def\gsim{\mathrel{\rlap{\lower4pt\hbox{\hskip1pt$\sim$}}
    \raise1pt\hbox{$>$}}}                % greater than or approx. symbol

\shorttitle{Record-breaking Storm Activity on Uranus in 2014}
\shortauthors{de Pater et al.}
\begin{document}
\title{Record-breaking Storm Activity on Uranus in 2014}

\author{Imke de Pater\altaffilmark{1,2,3}, L. A. Sromovsky\altaffilmark{4}, P. M. Fry\altaffilmark{4}, Heidi B. Hammel\altaffilmark{5}, Christoph Baranec\altaffilmark{6}, and Kunio Sayanagi\altaffilmark{7}}

\altaffiltext{1}{Astronomy Department, 501 Campbell Hall, University of California, Berkeley, CA 94720, USA; imke@berkeley.edu}
\altaffiltext{2}{Faculty of Aerospace Engineering, Delft University of Technology, NL-2629 HS Delft, The Netherlands}
\altaffiltext{3}{SRON Netherlands Institute for Space Research, 3584 CA Utrecht, The Netherlands}
\altaffiltext{4}{Space Science and Engineering Center, University of Wisconsin-Madison, Madison, WI 53706, USA}
\altaffiltext{5}{Association of Universities for Research in Astronomy, 1212 New York Avenue NW, Suite 450, Washington, DC 20005, USA}
\altaffiltext{6}{Institute for Astronomy, University of Hawai`i at M\={a}noa, Hilo, HI 96720-2700, USA}
\altaffiltext{7}{Department of Atmospheric and Planetary Sciences, Hampton University, Hampton, VA, 23668, USA}

\begin{abstract}
In spite of an expected decline in convective activity following the 2007 equinox of Uranus, eight sizable storms were detected on the planet with the near-infrared camera NIRC2, coupled to the adaptive optics system, on the 10-m W. M. Keck telescope on UT 5 and 6 August 2014. All storms were on Uranus's northern hemisphere, including the brightest storm ever seen in this planet at 2.2 $\mu$m, reflecting 30\% as much light as the rest of the planet at this wavelength. The storm was at a planetocentric latitude of $\sim$15$^{\circ}$N and reached altitudes of $\sim$330 mbar, well above the regular uppermost cloud layer (methane-ice) in the atmosphere. A cloud feature at a latitude of 32$^{\circ}$N, that was deeper in the atmosphere (near $\sim$2 bar), was later seen by amateur astronomers. We also present images returned from our HST ToO program, that shows both of these cloud features. We further report the first detection of a long-awaited haze over the north polar region. 
\end{abstract}


\maketitle

\def \Kepler {\textit{Kepler}}
\section{Introduction}

Seasonal forcing has explained hemispherical asymmetries on Earth \citep{Zachos01}, Mars \citep{Nakamura02}, Jupiter \citep{Orton94}, Saturn \citep{Greathouse05, Moses05, Orton05}, and Neptune \citep{Hammel07}. Uranus, with its pole almost in the ecliptic plane, provides a planetary obliquity extremum. Because it also lacks a measurable internal heat source, its weather depends more on solar energy than that of the other giant planets.  Hence, if seasonal forcing is important, it should be most apparent on Uranus. We report here extreme dynamic activity on Uranus that does not fit any existing seasonal model of this planet.

During the Voyager encounter with Uranus in 1986, only a handful of dim clouds were seen in its atmosphere. The typical contrast of these clouds (i.e., enhancement in albedo relative to the background) varied from $<1\%$ in the violet to $\sim7\%$ in the red \citep{Karkoschka98}. All those features were in its southern hemisphere, the hemisphere facing the Sun (and Earth) at that time. When Uranus's northern hemisphere came into view, HST images revealed an abundance of small cloud features at latitudes that had been in shadow for $\sim$40 years. The contrast of these features was highest at 1.6-1.8 $\mu$m, where it reached values over 2 orders of magnitude higher than the $\sim1\%$ seen by Voyager at visible wavelengths. The observed contrast is a function of atmospheric opacity and cloud reflectivity. At UV and visible wavelengths, gaseous opacity is dominated by Rayleigh scattering, while methane absorption dominates in the near-infrared. At 2.2 $\mu$m absorption by CH$_4$ and H$_2$ is so large that only the highest levels in Uranus's atmosphere are visible, as sunlight is absorbed deeper in the atmosphere. At 1.6-1.8 $\mu$m the opacity is small enough that pressure levels in the atmosphere down to several bar\footnotemark[8] are probed, the altitudes at which clouds typically form.

\footnotetext[8]{In MKS units, 1 bar = 10$^5$ Pascals}

When Uranus approached equinox in 2007 (i.e., when the Sun was shining directly on the equator), cloud activity at near-infrared wavelengths (1-2.2 $\mu$m) increased compared to earlier images at these same wavelengths. In particular, in 2004 one cloud feature in the southern hemisphere at $\sim34^{\circ}$S, known as the ``Berg," developed into a powerful storm\footnotemark[9] rising up to pressures of $\sim$0.6 bar; at 1.6 $\mu$m it was seen as a large morphologically interesting cloud system, while the top of the cloud core that reached up to these high altitudes was visible as a tiny (unresolved) feature at 2.2 $\mu$m \citep{Hammel05}. At 1.6 $\mu$m it became more elongated in longitude in 2007, at times displaying several small cloud features at 2.2 $\mu$m \citep{Sromovsky09, dePater11}. In 2005 this cloud complex started to migrate towards the equator \citep{Sromovsky09}, and it disintegrated in 2009 \citep{dePater11}.

\footnotetext[9]{We use the word ``storm" to refer to extremely bright (relative to the background atmosphere) and often large cloud features.}

The northern hemisphere, which has been rotating into view since the 1990s, contained a greater number of bright cloud features than seen during previous decades in the southern hemisphere, with the brightest long-lived feature at $\sim30^{\circ}$N, usually a double spot known as the ``Bright Northern Complex." This feature brightened considerably at times, and was usually also visible at 2.2 $\mu$m. At the peak of its brightness, the feature's cloud-top reached altitudes up to near 300 mbar \citep{Sromovsky07}. 
The brightest discrete features on Uranus have morphologies -- long extended sweeps of clouds and multiple features traveling together as a group --  that are suggestive of  underlying vortex systems deeper in the atmosphere. 

We have periodically imaged Uranus at high spatial resolution since its 2007 equinox to track the seasonal evolution of its atmosphere (this is the first post-equinoctial season observed with modern astronomical technology). The cloud morphology on Uranus has not seen many changes, with occasional small cloud features in addition to the Bright Northern Complex. While the south pole has been covered in a featureless haze since our first good views from Voyager 2, when the north pole came into view it was peppered with small discrete cloud features \citep{Sromovsky12b}. This difference may be a seasonal effect: the south pole has been seen during its summer and fall period, while the north pole has only been seen during springtime. Based on decades-long records of Uranus brightness \citep{Lockwood06}, we expect its north pole to become as hazy as its south pole \citep{Hammel07}, and perhaps cover the small cloud features.

\begin{figure*}
  \centering
  \resizebox{0.93\textwidth}{!}
   {
    \includegraphics{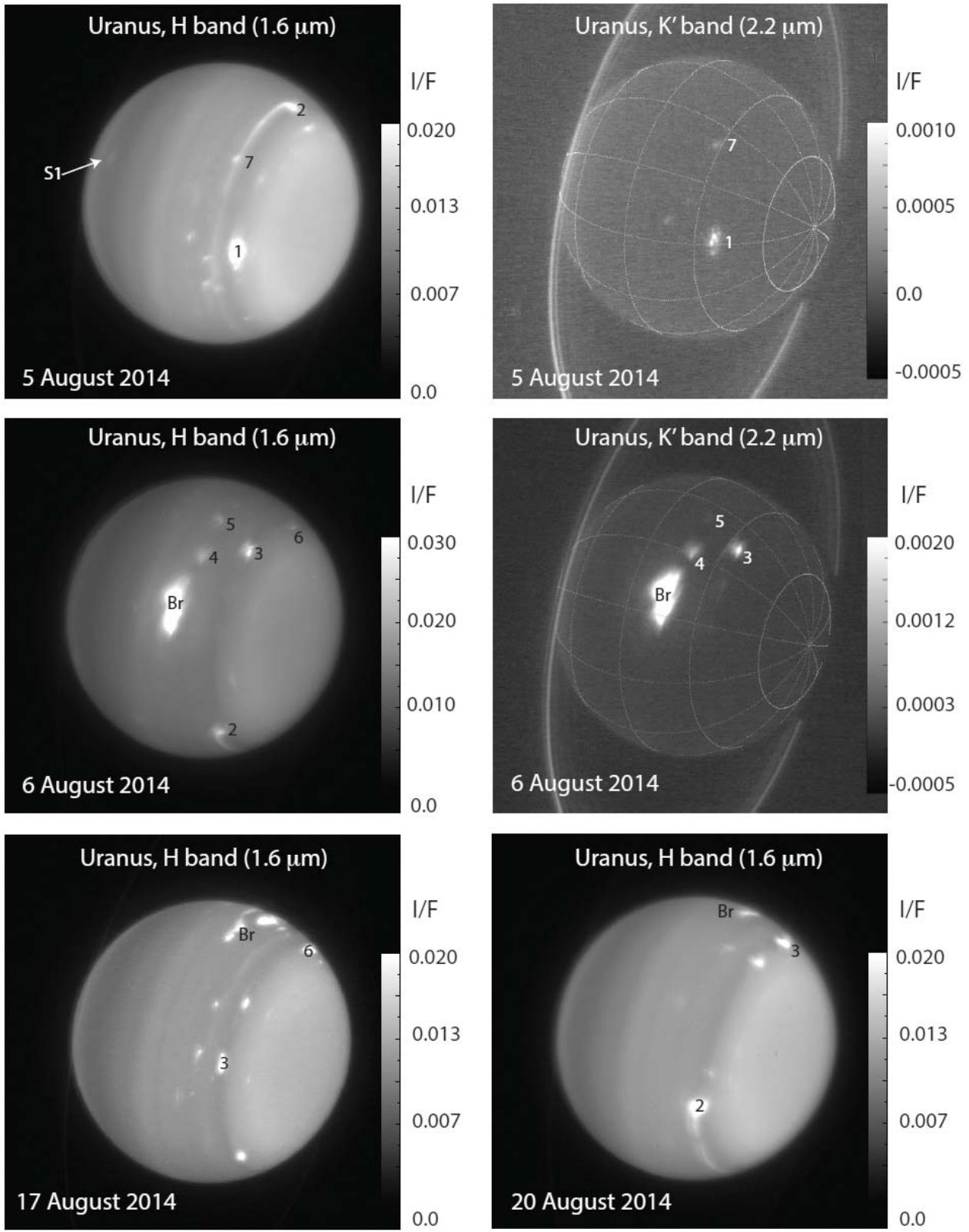}
   }
   \caption{Images of Uranus at H (1.6 $\mu$m) and K' band (2.2 $\mu$m) obtained with the 10-m Keck telescope on UT 5 and 6 August 2014 (top and middle row); the bottom row shows H band images taken on UT 17 and 20 August 2014. On 5 August the images were both obtained near UT 12:30.  On 6 August they were obtained at UT 15:29 (H) and UT 15:32 (K'). Note the extremely large storm system (labeled Br) on 6 August. Seven other features that are discussed in the paper are indicated as well. A feature seen in H band on the southern hemisphere on 5 August is marked with an arrow and the symbol S1. Numerous other features are also visible in the H band images when suitably enhanced. A longitude-latitude (planetocentric) grid has been superposed on the K' band images with a grid interval of 30$^{\circ}$.\label{fig:one}}

\end{figure*}

\section{Observations and Data Reduction}
\label{sec:obs}

Observations of Uranus were obtained on UT 5 and 6 August 2014 with the 10-m W. M. Keck II telescope on Maunakea (Hawaii), using the near-infrared camera NIRC2, coupled to the adaptive optics (AO) system \citep{Wizinowich00}. NIRC2 is a 1024$\times$1024 Aladdin-3 InSb array, which we used in its highest angular resolution mode, i.e., the NARROW camera at 9.94$\pm$0.03 mas per pixel \citep{dePater06}, which translates roughly to 140 km/pixel on these dates. Because the AO system was operating in a sub-optimal fashion, we did not achieve the expected (diffraction-limited) resolution.

Data were taken in the broadband H (1.48-1.78 $\mu$m) and K' (1.95-2.30 $\mu$m) filters, and the narrowband CH$_4$S (1.53-1.66 $\mu$m) filter, over a 4.5-hr period (11 – 15:30 UT) on each night. One additional image in each H and J (1.17-1.33 $\mu$m) band was obtained on UT 17 August at $\sim$15:10 and in H and K' band on UT 20 August 2014 at $\sim$13:40.  All images were processed using standard near-infrared data reduction techniques (flat-fielded, sky-subtracted, with bad pixels replaced by the median of surrounding pixels). We corrected geometric distortion using ``dewarp" routines provided by Brian Cameron of the California Institute of Technology\footnotemark[10]. Photometric calibrations were performed using the star HD1160 \citep{Elias82}.  We converted the observed flux densities to the dimensionless parameter I/F as in \citet{Hammel89}.

\footnotetext[10]{\url{http://www2.keck.hawaii.edu/inst/nirc2/forReDoc/post\_obs
erving/dewarp/nirc2dewarp.pro}}

Because Uranus has not exhibited notable atmospheric events over the past several years, we were surprised that the northern hemisphere of Uranus was unusually active on UT 5 and 6 August, and the activity continued throughout at least UT 20 August (Fig.~\ref{fig:one}). A total of about eight sizable cloud features (labeled on the images) were seen in the northern hemisphere in each of the broadband H and K'  filters on UT 5 and 6 August. Assuming the features would follow \citet{Sromovsky09} 13-term Legendre polynomial fit to Uranus's wind profile, we were able to identify a few of these features on UT 17 and 20 August, as indicated. However, without continuous time coverage we cannot be absolutely sure these are the same features.

Table~\ref{tab.1} summarizes the latitudes of these eight features. Features labeled 2 and 3 are in a latitude range that makes them possible candidates to be remnants of the Bright Northern Complex \citep{Sromovsky07}. Zonal banding is seen throughout the atmosphere, with the northern hemisphere being ``hazier" than the south. For the first time, we saw a thin haze layer forming over the north pole. 

In the H band, numerous smaller cloud features were visible as well. On UT 5 August, some clouds (Feature 2, as well as some smaller clouds) show what look like narrow plumes of cloud particles trailing to the east (down in the picture) at 1.6 $\mu$m. On this date, a small cloud can be seen in the southern hemisphere as well. For altitude determination, the images captured on UT 5 and 6 August were complemented with images taken through a narrowband CH$_4$S  filter; visually, these images look very similar to the H band images, but probe somewhat different atmospheric depths.

\begin{figure*}
  \centering
  \resizebox{0.85\textwidth}{!}
   {
    \includegraphics{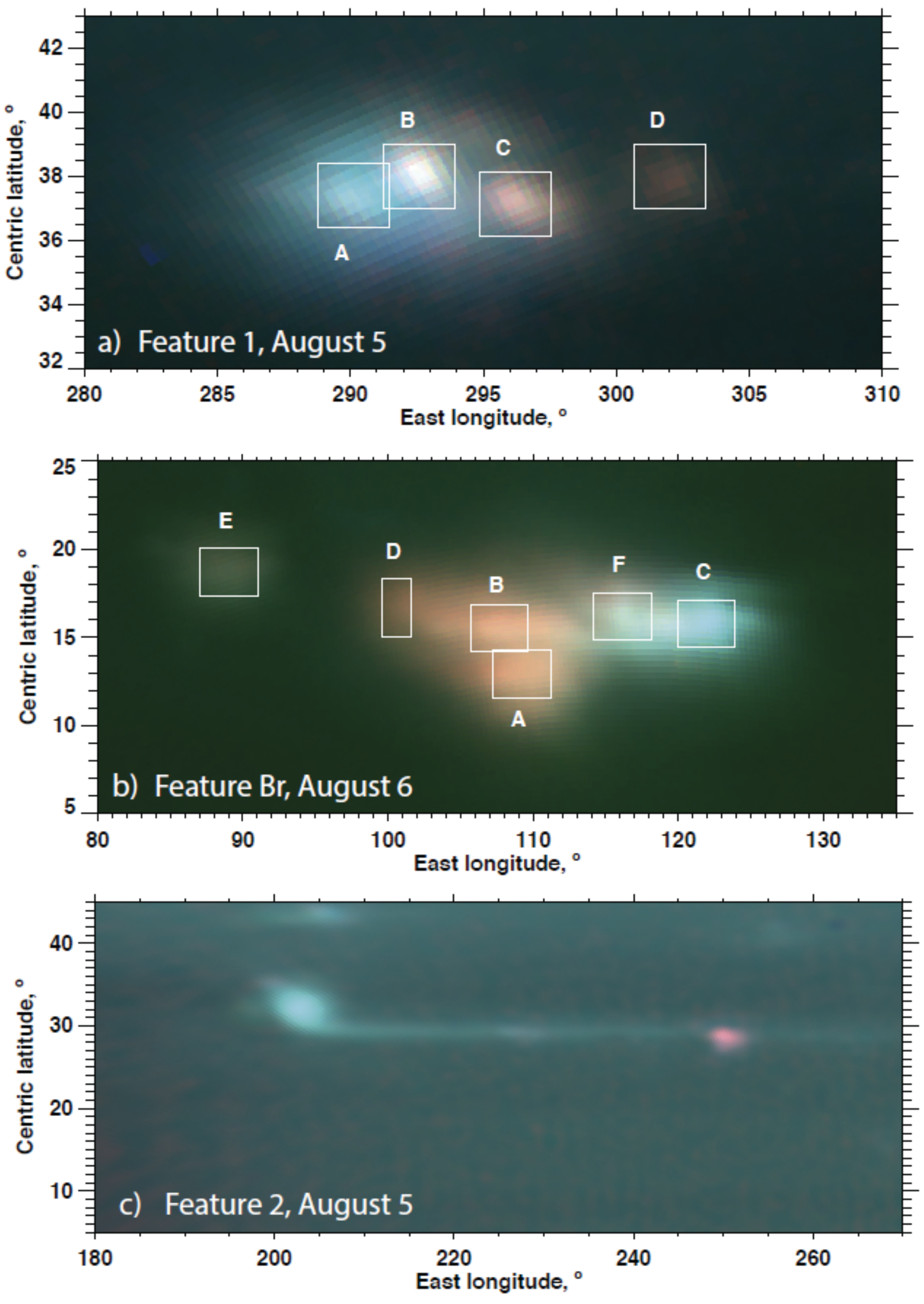}
   }
   \caption{Color composite images of Features 1, 2 and Br, projected on a rectangular grid. The K' band image, sensitive to the highest altitudes, is shown in red; H band, sensitive to the deepest levels, is shown in blue, and CH$_4$S in green. The images of Feature 1 were taken on UT 5 August 2014 near 14:30, Feature 2 on UT 5 August 2014 near 11:30, and of Br on UT 6 August 2014 near 14:20. The boxes A-D for Feature 1, and A-F for Br indicate the areas where cloud altitudes were determined (Table~\ref{tab.1}). Feature E is the same as Feature 4 in Fig. 1.\label{fig:two}}

\end{figure*}

\begin{figure}
  \centering
  \resizebox{1.0\columnwidth}{!}
   {
    \includegraphics{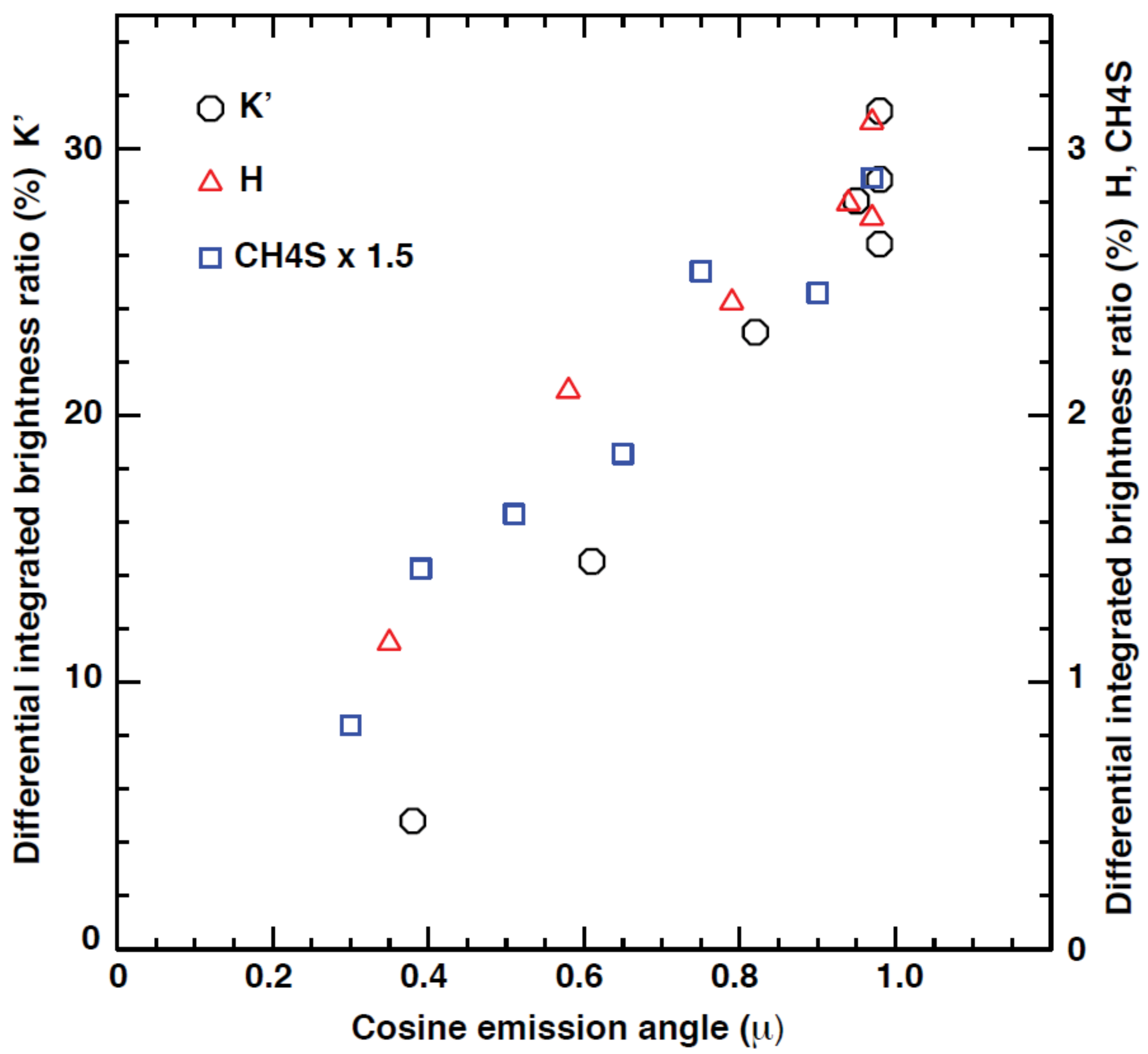}
   }
   \caption{The ratio (expressed as a percent) of the differential integrated brightness of the cloud as a function of the emission angle, $\mu$. The intensity scale for K' band is given on the left; for H and CH$_4$S bands on the right (the I/F for the CH$_4$S band was multiplied by 1.5). Note that the cloud in K' band reaches 30\% of the total reflected light from the rest of the planet.\label{fig:three}}
\end{figure}

\section{Discussion}
\label{sec:iq}

The brightness and morphology of Uranus's cloud activity on UT 6 August is unprecedented in over a decade of 2.2-$\mu$m imaging with Keck. Figure~\ref{fig:two} shows Features 1, 2 (UT 5 August) and Br (UT 6 August) as color composites (K' = red; CH$_4$S = green; H = blue) after projection onto a rectangular latitude-longitude grid. Feature 1 is composed of several compact clouds, visible at both 1.6 and 2.2 $\mu$m, extending almost $\sim15^{\circ}$ in longitude and $\sim3^{\circ}$ in latitude, while Feature Br extends over $\sim25^{\circ}$ in longitude, i.e., almost 10,000 km in extent, and $\sim7^{\circ}$ in latitude. Interestingly, the west side of Br is much redder (i.e., relatively brighter at 2.2 microns, and thus at higher altitudes in the atmosphere) than the east side, indicative of a difference in cloud altitude. Feature 1 also shows such a gradient, but here the east side is reddest. Since the winds blow in opposite directions at these two latitudes (e.g., \citealt{Sromovsky12b}), both features have in common that the clouds in the down-wind direction are higher in the atmosphere than in the up-wind direction, which perhaps may indicate the effect of vertical wind shear on a convective updraft. 

Figure 2c shows a color composite of Feature 2. This image has been processed both to enhance the contrast and take out artifacts induced by the AO system working in a sub-optimal fashion. The double-lobed structure induced by the latter is seen particularly well in the trail of Feature 2 (Fig.~\ref{fig:one}). We deconvolved the images by using a double Gaussian PSF with offset of 6 pixels and relative amplitude of 0.4 for the secondary lobe. Eight consecutive H band images were processed this way, as well as an image close in time in the K' and CH$_4$S bands. Following \citet{Fry12}, all images were projected on a rectangular latitude-longitude grid, and then combined. Feature 2 is moving to the west and faster than the atmosphere at latitudes just to the south.  If it injected cloud material to its south, it would indeed make the long streak we see trailing to the east. The meridional wind shear at this latitude is about -0.08 degree of longitude per hour per degree of latitude.  Assuming the streamer is 3$^{\circ}$ S of the storm center, it would extend eastward (relative to the storm) at a rate of 0.24$^{\circ}$ hour$^{-1}$ or 5.8$^{\circ}$ day$^{-1}$, and thus take about 10 days to extend backward by $\sim$60$^{\circ}$. The morphology of this cloud, with its long slightly slanted tail ($\sim$1$^{\circ}$ in latitude over $\sim$60$^{\circ}$ in longitude towards the east) reminds us of the early morphology of the 2010 storm on Saturn, that developed into a storm system encompassing the entire circumference of the planet \citep{Sanchez11, Sanchez12, Garcia-Melendo13, Sayanagi13}. 

Features 1 and Br are both relatively compact and accompany cloudless areas around them; such morphologies suggest that these features might be coupled to vortex systems deeper in the atmosphere. The clouds likely form when gas rising in the atmosphere becomes cold enough for particular constituents, such as methane gas, to condense and form clouds, similar to the orographic clouds associated with dark spots on Neptune, which have been modeled numerically \citep{Stratman01}. As shown, by UT 17 August, the storm's brightness had decreased substantially. If indeed these features are orographic clouds, drastic fluctuations in brightness would not be surprising.

\begin{figure}
  \centering
  \resizebox{1.0\columnwidth}{!}
   {
    \includegraphics{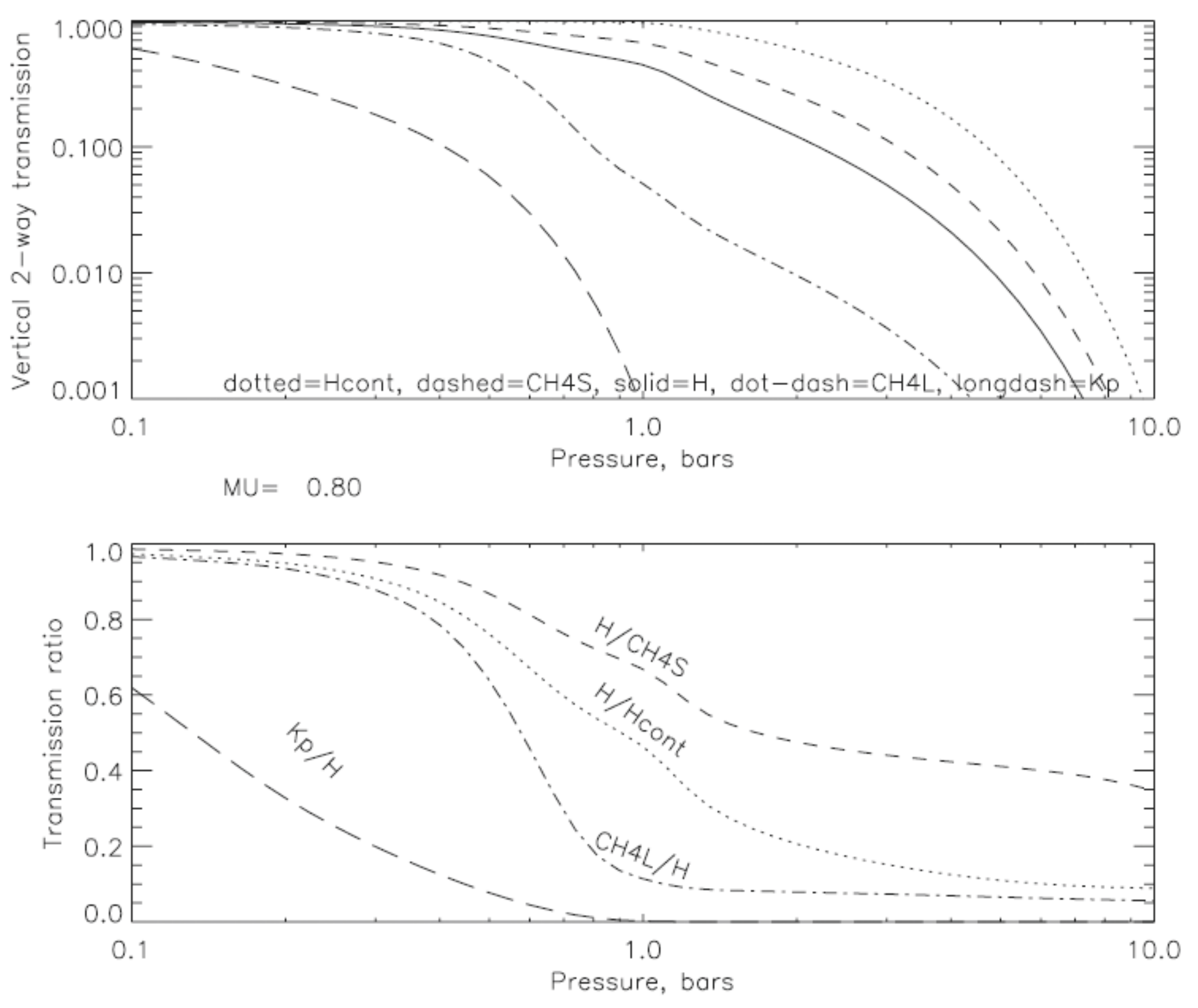}
   }
   \caption{Models of the vertical 2-way transmission of several filters (top) and the transmission ratios (bottom) for a zenith angle of 0.8. In a clear atmosphere, the observed I/F should follow the 2-way transmission. Such models are used to determine the altitudes of the various features, using brightness ratios in different filters, following \citet{dePater11} and \citet{Sromovsky12a}.\label{fig:four}}

\end{figure}

\begin{figure*}
  \centering
  \resizebox{0.95\textwidth}{!}
   {
    \includegraphics{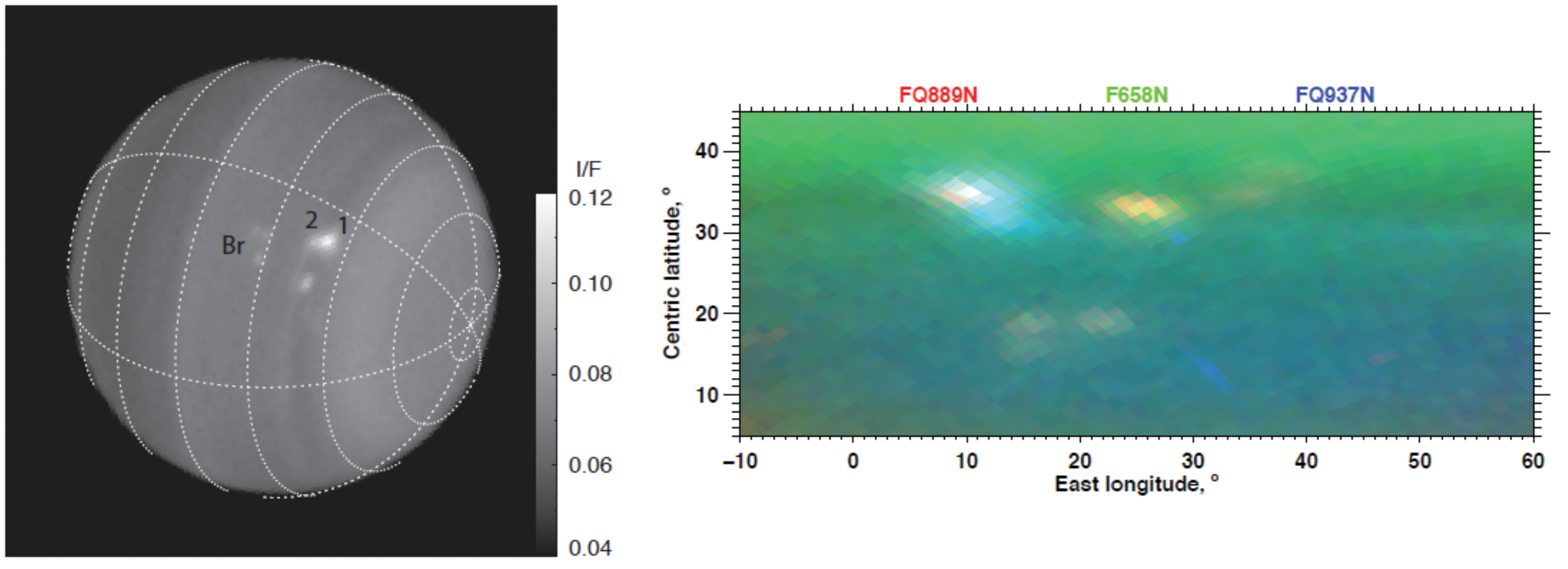}
   }
   \caption{a) HST image taken on UT 14 October 2014 in the 845M filter.  b) 3-color composite after projection on a rectangular latitude-longitude grid. The colors and filters are indicated at the top of the image (FQ889N in red; Q658N in green; FQ937N in blue). As in Fig. 2, in this composite, high altitude features are red and deep features are blue.\label{fig:five}}

\end{figure*}

Rectilinear projections of the images from UT 6, 17 and 20 August show that the morphology of the feature had changed so much that it was not possible to make unambiguous measurements of the feature's drift rate. Using local bright components of the feature on each date we found two possible drift rates: 0.183$^{\circ}\pm0.001^{\circ}$ hour$^{-1}$ or 0.148$^{\circ}\pm0.001^{\circ}$ hour$^{-1}$ eastward, The former is more probable as it is very similar to the 0.18$^{\circ}$ hour$^{-1}$ drift rate predicted for a latitude of 15$^{\circ}$N based on \citet{Sromovsky09} 13-term Legendre polynomial fit to Uranus's wind profile. 

The differential integrated brightness was obtained by integrating the brightness (in units of reflectivity, I/F, above the background level) in each pixel of the cloud. We then took the ratio of this number to the total reflectivity of a Uranus without the presence of discrete cloud features. In order to determine the reflectivity of Uranus without clouds, we constructed an image from all data combined. By using a combination of median averaging and removal of excess cloud brightness above the background we were able to construct an image of Uranus without clouds. The advantage of using this method, rather than using images taken on different nights in previous years that showed no cloud features, is that it keeps the same viewing geometry and photometric calibration. Hence these numbers do not suffer from photometric calibration errors (which typically are of order 10-15\%). 

In the absence of gas opacity, the brightness of a perfectly reflecting cloud of finite dimensions would vary with $\mu$ $\mu_0$ ($\mu$ and $\mu_0$ are cosines of the emission and incidence angles), due to foreshortening of the projected area both with respect to the Sun and Earth. Increased gas opacity near the limb (i.e, methane and hydrogen absorption, which is strongest in K' band) would enhance the effect (the path length, and hence one-way opacity increases with 1/$\mu$). An optically thin cloud, on the other hand, would brighten near the limb (with 1/$\mu$) due to an increase in optical depth, $\tau$, of cloud particles along the line of sight (until $\tau\sim0.15$; see \citealt{Dunn10}). We found a near-linear increase in Br's brightness as a function of $\mu$ (Fig.~\ref{fig:three}), suggesting that the clouds must be optically thin. 

The most remarkable result is the cloud's total brightness in K' band: it represents $\sim$30\% of the total light reflected by the rest of the planet at this wavelength. This is $\sim$2 times larger than what was seen in 2005 for the ``brightest cloud feature ever observed" on Uranus \citep{Sromovsky07}. The integrated brightness ratio in H band is a factor of 10 lower than that in K' band, $\sim$3\%. This is somewhat less than was observed at H in 2005, which suggests that a much larger fraction of the 2014 cloud resides at very high levels in Uranus's atmosphere.

By using the ratio of the cloud excess reflectivity in the K' vs. H bands, and in the H vs. the CH$_4$S bands, we can determine an effective pressure (i.e., altitude) of the various cloud features. To obtain a rough estimate, we calculate the ratio of the integrated I/F values relative to the background (Table~\ref{tab.1}) and compare this with calculations of ratios for reflecting layers at different pressure levels, as shown in Figure~\ref{fig:four} \citep{dePater11, Sromovsky12a}. The pressures tabulated for the sub-elements of larger features in Table 1 were derived with the same model using a linear-regression fit \citep{Sromovsky12a}, where for each point in the boxes in Figure 2 the I/F values in the K' band were plotted against those in the H band, and similarly in the H vs. CH$_4$S band. 

As expected from the numerous clouds detectable in the K' band, most features are at relatively high altitudes. Feature 2, which was not detected at K', is the deepest feature, located at altitudes below the 1-bar pressure level. All other features are at altitudes above the methane condensation level at 1.2 bar \citep{Lindal87}, and hence we suggest that these high clouds are composed of methane ice. 

Interestingly, some parts of the bright feature Br reach altitudes up to $\sim$330 mbar, whereas other parts are much deeper in the atmosphere, near the 700-mbar level. The fact that such a large cloud is visible at such high altitudes suggests that there must have been strong updrafts or waves at or (shortly) before the time of observation. This result, combined with the observation of so many clouds at high altitudes during these observations, suggests that Uranus was going through a period of extreme dynamic activity. 

In order to learn more about this unusual activity, we shared our data immediately with the amateur community, and issued an alert through the International Outer Planets Watch. The amateur community responded with an extensive observation campaign in close coordination with our team. Throughout the 2014 fall season they observed Uranus. By early October several amateur astronomers had reported the detection of a bright cloud feature on the planet's disk , using telescopes varying in size from 14-inch up to the 1-m telescope at Pic-du-Midi and broadband filters spanning $\sim$650-850 nm range. The cloud detected by them, however, was not the Feature Br seen on UT 6 August; it was cloud Feature 2 seen on UT 5, 6  and 20 August (Fig.~\ref{fig:one}), the only cloud that was much deeper in the atmosphere, and that displayed a tail reminiscent of the early morphology of the 2010 storm on Saturn. 

These amateur observations were used to trigger our Target of Opportunity (ToO) program on HST (GO13712, PI: K. Sayanagi). These observations were carried out on UT 14 October 2014; images are shown in Figure~\ref{fig:five}. An image taken at 845 nm (845M band), where the contrast of features on Uranus's disk is large, is shown in Panel a. Panel b shows a 3-color composite of a portion of the image after projection on a rectangular latitude-longitude grid: The highest levels in the atmosphere are probed in the FQ889N filter (red); the deepest levels in the atmosphere are probed in the FQ937N filter (blue); the Q658N filter probes intermediate altitudes (green).  

Feature 2, with its extended streamer, is visible just north of the center of the disk at a latitude of 32$^{\circ}$N. At this time it looks like a complex feature, visible in all filters, i.e., it appears to extend from the deep atmosphere to the highest layers in the atmosphere. Rather than just Feature 2, however, we see instead the chance appearance of both Features 1 and 2 at nearly the same longitude; Feature 1, being higher in the atmosphere (Fig.~\ref{fig:one}), gives the complex feature the red color near 35$^{\circ}$N. Feature 2's extended streamer is visible only in the blue (FQ937N filter). 

The extended tail on the south side shows a slightly steeper incline, by $\sim$1$^{\circ}$ in latitude over $\sim$40$^{\circ}$ of longitude. The triplet of features near 17$^{\circ}$N is most likely connected to feature Br seen on UT 6 August (and perhaps on the limb on UT 17 and 20 August). It is visible at high levels in the atmosphere. These HST and amateur images will be analyzed in detail in a future paper.

Neither our Keck or HST observations revealed a dark spot. Such spots have been seen in the past \citep{Hammel09, Sromovsky12a}, and may be a signature of a vortex system deeper in the atmosphere. The non-detection in our Keck images is not surprising, as such detections require an excellent AO correction. We also did not detect a dark spot in the HST ToO images. Note, though, that not detecting a dark spot does not imply that there is no vortex system.

Circulation models predict the largest hemispheric asymmetry at equinox, and the most significant convective events in the late winter \citep{Sussman12}, i.e., they would be visible when the region first comes into view after a long cold winter, as the cloud features that were seen in the 1990s with HST, as mentioned in the introduction \citep{Karkoschka98}. These models thus predict that cloud activity most likely would subside after equinox, the opposite of what we report here. Since 2007, Uranus's cloud activity had seemed to decrease slightly, making these new extremely bright cloud and the many other features seen in K' band even more surprising.

\subsection{Conclusions}

Observations of Uranus, taken with the NIRC2 camera on the 10-m W. M. Keck telescope in August 2014 revealed an unusually dynamic planet. In particular, we detected the brightest feature ever detected in K' band, reflecting 30\% as much light as the rest of the planet at that wavelength. The feature is optically thin, and its cloud tops extend over pressure levels varying from $\sim$700 mbar up to 330 mbar. The cloud that was at the deepest levels in the atmosphere, $\sim$2 bar, was later seen by amateur astronomers. Both clouds were imaged with HST on UT 14 October 2014; the bright feature showed a considerable evolution in morphology. Our Keck observations also showed the development of the long-awaited haze over Uranus's north polar cap.

These unexpected observations remind us keenly of how little we understand about dynamics in the atmospheres of the outer planets in our Solar System. Planets around other stars are now known to be extremely common (e.g., \citealt{Petigura13}). The Kepler mission has revolutionized our knowledge of such exoplanets, with the detection of over 4000 planet candidates and nearly 1000 confirmed planets, $\sim$6\% of which are Uranus-sized (radii between 3.5 and 4.5 Earth radii), and many more are even smaller \citep{Batalha13}. Increasingly, the atmospheres of transiting exoplanets are being characterized \citep{Nikolov14, Swain14, Ranjan14} using our own planets as templates. Yet, many aspects of our own giant planets remain elusive. It is clear from this paper that a full understanding of the dynamic activity in Uranus's atmosphere requires frequent observations of this planet at high spatial resolution throughout the changing seasons. Ideally, observations would be obtained simultaneously at multiple wavelengths, from the UV into the radio regime, enabling probing levels in the atmosphere from the stratosphere down to tens of bars in the deep troposphere. But even with such data, some questions may ultimately only be answered with a Uranus flyby or orbiter.

\acknowledgments

The W. M. Keck Observatory is operated as a scientific partnership among the California Institute of Technology, the University of California, and the National Aeronautics and Space Administration and was built with financial support of the W. M. Keck Foundation. L. S. and P. F. acknowledge support from NASA Planetary Astronomy Grant NNX13AH65G. C. B. acknowledges support from the Alfred P. Sloan Foundation. K. M. S. is the PI of HST Grant GO13712.  K. M. S. acknowledges support from NASA Planetary Atmospheres Grant NNX14AK07G and NSF AAG Grant 1212216. The authors wish to recognize and acknowledge the very significant cultural role and reverence that the summit of Maunakea has always had within the indigenous Hawaiian community. We are most fortunate to have the opportunity to conduct observations from this mountain.

{\it Facility:} \facility{Keck:II (NIRC2-NGS), HST}

\bibliographystyle{apj.bst}

\newpage

\setcounter{table}{0}
\small

\small
\begin{sidewaystable}[!ht]  %\centering
\small
\vspace{0.0in}
\caption{Uranus cloud properties on UT 5 and 6 August, 2014}
\vspace{0.15in}
\begin{tabular}{l c c c c c c c c c}
%% |l|l| to left justify each column entry
%% |c|c| to center each column entry
%% use of \rule[]{}{} below opens up each row
\hline
\hline
Feature &Pc.Latitude$^1$ &Pg.Latitude$^1$& Time$^2$&E.Longitude$^2$& $\mu^2$  & Total I/F$^3$   & Total I/F$^3$  & Total
I/F$^3$ &Pressure$^4$ \\
  & ($^\circ$)  & ($^\circ$)  & (month/date:hr:min--hr:min) & ($^\circ$) &   &Kp-band &H-band &CH$_4$S-band &  (mbar) \\
\hline
1&37.8$\pm$0.4&39.1$\pm$0.4&8/5:11:14--13:40&297&0.71-0.98&0.35&6.8&12.2&400 -- 1200  \\
1A&37.4&38.7&8/5:11:14--11:47&290.1&0.71-0.81& & & &720$\pm$40\\
1B&38.0&39.3&8/5:11:14--11:47&292.6&0.71-0.81& & & &680$\pm$90\\
1C&37.1&38.4&8/5:11:14--11:47&296.2&0.71-0.81& & & &620$\pm$80\\
1D&38.0&39.3&8/5:11:14--11:47&302.0&0.71-0.81& & & &420$\pm$50\\
2&32.1$\pm$0.5&33.3$\pm$0.5&8/5:11:14--11:35&201.9&0.73--0.79&--&3.8&8.2&$\sim$~2000\\
3&34.3$\pm$0.6&35.5$\pm$0.6&8/6:13:18--13:33&76&0.99&0.66&5.1&4.8&460$\pm$90 \\
4&19.0$\pm$0.5&19.8$\pm$0.5&8/6:13:18--13:33&89&0.95--0.99&0.36&1.6&1.9&420$\pm$100  \\
5&16.3$\pm$0.7&17.0$\pm$0.7&8/6:13:18--13:33&68&0.97--0.98&0.087&1.9&2.3&600$\pm$100  \\
6&36.5$\pm$0.5&37.8$\pm$0.5&8/6:13:18--13:33&37&0.83--0.88&0.075&2.9&1.0& 550$\pm$100 \\
7&28.1$\pm$0.3&29.2$\pm$0.3&8/5:13:18--13:33&250&0.73--0.99&0.09&1.1&--&550$\pm$100 \\
Br&14.8$\pm$0.7&15.5$\pm$0.7&8/6:14:11--15:15&113&0.90-0.99 &6.2&44&51&300 -- 700\\
BrA&12.93&13.52&8/6:14:11--14:27&109.2&0.90-0.95& & & &410$\pm$80\\
BrB&15.53&16.23&8/6:14:11--14:27&107.7&0.90-0.95& & & &370$\pm$20\\
BrC&15.80&16.51&8/6:14:11--14:27&121.9&0.90-0.95& & & &675$\pm$25\\
BrD&16.67&17.41&8/6:14:11--14:27&100.6&0.90-0.95& & & &330$\pm$60\\
BrE&18.73&19.55&8/6:14:11--14:27&89.1&0.90-0.95& & & &420$\pm$100\\
BrF&16.20&16.93&8/6:14:11--14:27&116.2&0.90-0.95& & & &620$\pm$50\\
\hline
\end{tabular}\label{tab.1}

$^1$: Planetocentric (Pc) and Planetographic (Pg) Latitude. Numbers with errorbars are averaged over all
data. Values without errorbars (1A--D; BrA--F) correspond to the centers of the boxes
in Fig. 2.

$^2$: The UT time range, East-longitude and approximate emission angle
corresponding to the pressures quoted for the feature. 

$^3$: Integrated I/F above the background level, i.e., we summed the
I/F (minus background) in each pixel of the cloud. The integrated I/F
for the entire planet, without the discrete cloud features, was 19.6
in the K', 1420 in the H, and 2638 in the CH$_4$S bands.

$^4$: Pressure levels: pressures were derived from linear regression fits
to the Kp/H and H/CH$_4$S data, and the integrated I/F for the
eight features, as provided in the table.  

\end{sidewaystable}

\end{document}